\documentclass[a4paper]{jpconf}

\usepackage{graphicx}
\usepackage{cleveref}
\usepackage{feynmp-auto}
\usepackage{subcaption}
\usepackage{accents}
\usepackage{xspace}
\usepackage{enumitem}

\newcommand{\bg}{\ensuremath{\beta_g}\xspace}
\newcommand{\bgs}{\ensuremath{\beta_g^2}\xspace}
\newcommand{\MET}{\ensuremath{E_{T}^{miss}}\xspace}

\newcommand{\ggf}{$gg$F\xspace}

\begin{document}

\title{Anomalies in the production of multiple leptons at the LHC}

\author{Stefan von Buddenbrock$^{1}$ and Bruce Mellado$^{1,2}$}

\address{$^{1}$ School of Physics and Institute for Collider Particle Physics, University of the Witwatersrand, Johannesburg, Wits 2050, South Africa} 

\address{$^{2}$ iThemba LABS, National Research Foundation, PO Box 722, Somerset West 7129, South Africa}

\ead{stef.von.b@cern.ch,bmellado@mail.cern.ch}

\begin{abstract}
Based on a number of features from proton-proton collisions taken during Run 1 data taking period at the LHC, a boson with a mass around the Electro-Weak scale was postulated such that a significant fraction of its decays would comprise the Standard Model (SM) Higgs boson and an additional scalar, $S$. One of the phenomenological implications of a simplified model, where $S$ is treated a SM Higgs boson, is the anomalous production of high transverse momentum leptons. A combined study of Run 1 and Run 2 data is indicative of very significant discrepancies between the data and SM Monte Carlos in a variety of final states involving multiple leptons with and without $b$-quarks. These discrepancies appear in corners of the phase-space where different SM processes dominate, indicating that the potential mismodeling of a particular SM process is unlikely to explain them. Systematic uncertainties from the prediction of SM processes evaluated with currently available tools seem unable to explain away these discrepancies. The internal consistency of these anomalies and their interpretation in the framework of the original hypothesis is quantified. 
\end{abstract}

\section{Introduction}

An early study in 2015 considered the possibility of a heavy scalar, $H$, being compatible with several  LHC Run 1 measurements~\cite{vonBuddenbrock:2015ema}.
The result of this study had shown that with a single parameter \bgs (the scale factor for the production cross section of $H$) a set of ATLAS and CMS physics results could be fit with a significance of $3\sigma$. Using an effective vertex, the best fit mass of $H$ was found to be at $m_H=272^{+12}_{-9}$\,GeV. This study included, but was not limited to, the production of multiple leptons in association with $b$-jets, as reported by the search for the SM Higgs boson in association with top quarks. Other multi-lepton final states predicted in Ref.~\cite{vonBuddenbrock:2016rmr} and verified in Refs.~\cite{vonBuddenbrock:2017gvy,vonBuddenbrock:2019ajh}  were not included in the significance reported in Ref.~\cite{vonBuddenbrock:2015ema}. Now it seems evident that the multi-lepton final states reported in  Refs.~\cite{vonBuddenbrock:2017gvy,vonBuddenbrock:2019ajh} displayed sings of discrepancies with respect to SM predictions already in Run 1 data sets. Early discrepancies from Run 1 data sets not considered in  Ref.~\cite{vonBuddenbrock:2015ema} include opposite sign di-leptons and missing transverse energy with a full hadronic jet veto (see Ref.~\cite{vonBuddenbrock:2017gvy}) or di-leptons in association with at least one $b$-jet~\cite{Aad:2014mfk,Khachatryan:2016xws}, among others. 

Following a discussion of the results in Ref.~\cite{vonBuddenbrock:2015ema}, the next point of interest was to explore the possibility of introducing a scalar mediator $S$ (instead of using effective vertices), such that $H$ could decay to $Sh$, $SS$, and $hh$~\cite{vonBuddenbrock:2016rmr}.
The $S$ was assumed to have globally re-scaled Higgs-like couplings, such that its branching ratios (BRs) could be fixed. In this setup, and in the light of the results in Ref.~\cite{Aaboud:2017uak} where the 100\% branching ratio of $S$ into Dark Matter was ruled out, multi-lepton final states became a focus.
The possibility of embedding $H$ into a Type-II two Higgs doublet model (2HDM) was also discussed, where the allowed parameter space of the model was reported in Ref.~\cite{vonBuddenbrock:2016rmr,vonBuddenbrock:2018xar}.
More importantly, a predictive set of potential search channels for the new scalars was shown.
Several of these predictions were tested and expanded upon in Refs.~\cite{vonBuddenbrock:2017gvy,vonBuddenbrock:2019ajh}.

\section{Simplified Model}
\label{sec:model}

In terms of interactions, $H$ is assumed to be linked to electro-weak symmetry breaking in that it has Yukawa couplings and tree-level couplings the the weak vector bosons $V$ ($W^\pm$ and $Z$).
After electro-weak symmetry breaking, the Lagrangian describing $H$ is Higgs boson-like.
Omitting the terms that are irrelevant in this analysis, $H$ interacts with the SM particles in the following way:
\begin{equation}
\mathcal{L} \supset -\beta_{g}\frac{m_t}{v}t\bar{t}H + \beta_{_V}\frac{m_V^2}{v}g_{\mu\nu}~V^{\mu}V^{\nu}H.  \label{eqn:H_production}
\end{equation}
These are the the Higgs-like couplings for $H$ with the top quark ($t$) and the heavy vector bosons, respectively.
The strength of each of the couplings is controlled by a free parameter: $\beta_g$ for the $H$-$t$-$t$ interaction and $\beta_V$ for the $H$-$V$-$V$ interaction.
The omitted terms include the Yukawa couplings to the other SM fermions and self-interaction terms for $H$.
It can be expected that the couplings to the other SM fermions would also differ by a factor like \bg, however the effect would not make a noticeable difference to the analysis considered here and therefore these terms are neglected. The vacuum expectation value $v$ has a value of approximately $246$~GeV.

\begin{figure}
    \centering
    \begin{subfigure}[b]{\textwidth}
    \centering
    \begin{fmffile}{ggf}
	        \begin{fmfgraph*}(150,120)
	            \fmfstraight
	            \fmfleft{g1i,g2i}
	            \fmfright{p1,ho,p2,so,p3}
	            \fmf{gluon,tension=2}{g1i,g1t}
	            \fmf{gluon,tension=2}{g2i,g2t}
	            \fmf{fermion}{g2t,g1t}
	            \fmf{fermion}{g1t,ttH}
	            \fmf{fermion}{ttH,g2t}
	            \fmf{dashes,tension=1.5,label=$H$,l.side=left}{ttH,Hsh}
	            \fmf{dashes,tension=1.5}{Hsh,so}
    	        \fmf{dashes,tension=1.5}{Hsh,ho}
	            \fmflabel{$S$}{so}
	            \fmflabel{$h$}{ho}
	        \end{fmfgraph*}
	    \end{fmffile}
	    \caption{Gluon fusion (\ggf).}
	    \label{fig:ggf}
	\end{subfigure}
	
	\begin{subfigure}[b]{0.45\textwidth}
    \centering
    \begin{fmffile}{tth}
	        \begin{fmfgraph*}(120,150)
	            \fmfstraight
	            \fmfleft{p1,g1i,pm,g2i,p2}
	            \fmfright{pr1,t1o,ho,so,t2o,pr2}
	            \fmf{gluon,tension=1.5}{g1i,g1t}
	            \fmf{gluon,tension=1.5}{g2t,g2i}
	            \fmf{fermion}{t1o,g1t}
	            \fmf{fermion}{g1t,ttH}
	            \fmf{fermion}{ttH,g2t}
	            \fmf{fermion}{g2t,t2o}
	            \fmf{dashes,label=$H$,l.side=left}{ttH,Hsh}
	            \fmf{dashes}{Hsh,so}
    	        \fmf{dashes}{Hsh,ho}
	            \fmflabel{$S$}{so}
	            \fmflabel{$h$}{ho}
	            \fmflabel{$\bar{t}$}{t1o}
	            \fmflabel{$t$}{t2o}
	        \end{fmfgraph*}
	    \end{fmffile}
	    \caption{Top pair associated production ($ttH$).}
	    \label{fig:ttH}
	\end{subfigure}
	\hfill
	\begin{subfigure}[b]{0.45\textwidth}
    \centering
    \begin{fmffile}{th}
	        \begin{fmfgraph*}(120,150)
	            \fmfstraight
	            \fmfleft{p1,g1i,pm,g2i,p2}
	            \fmfright{pr1,t1o,ho,so,t2o,pr2}
	            \fmf{fermion}{g1i,g1t}
	            \fmf{fermion}{g2i,g2t}
	            \fmf{fermion}{ttH,t1o}
	            \fmf{plain,tension=2.4}{ttH,g1t}
	            \fmf{boson,label=$W^\pm$,l.side=left}{g1t,g2t}
	            \fmf{fermion}{g2t,t2o}
	            \fmf{dashes,label=$H$,l.side=left}{ttH,Hsh}
	            \fmf{dashes}{Hsh,so}
    	        \fmf{dashes}{Hsh,ho}
	            \fmflabel{$S$}{so}
	            \fmflabel{$h$}{ho}
	            \fmflabel{$t$}{t1o}
	            \fmflabel{$j$}{g2i}
	            \fmflabel{$b$}{g1i}
	            \fmflabel{$j^\prime$}{t2o}
	        \end{fmfgraph*}
	    \end{fmffile}
	    \caption{Single top associated production ($tH$).}
	    \label{fig:tH}
	\end{subfigure}
    \caption{The representative Feynman diagrams for the leading order production modes of $H$ and its subsequent decay to $Sh$.}
    \label{fig:feynman_diagrams}
\end{figure}
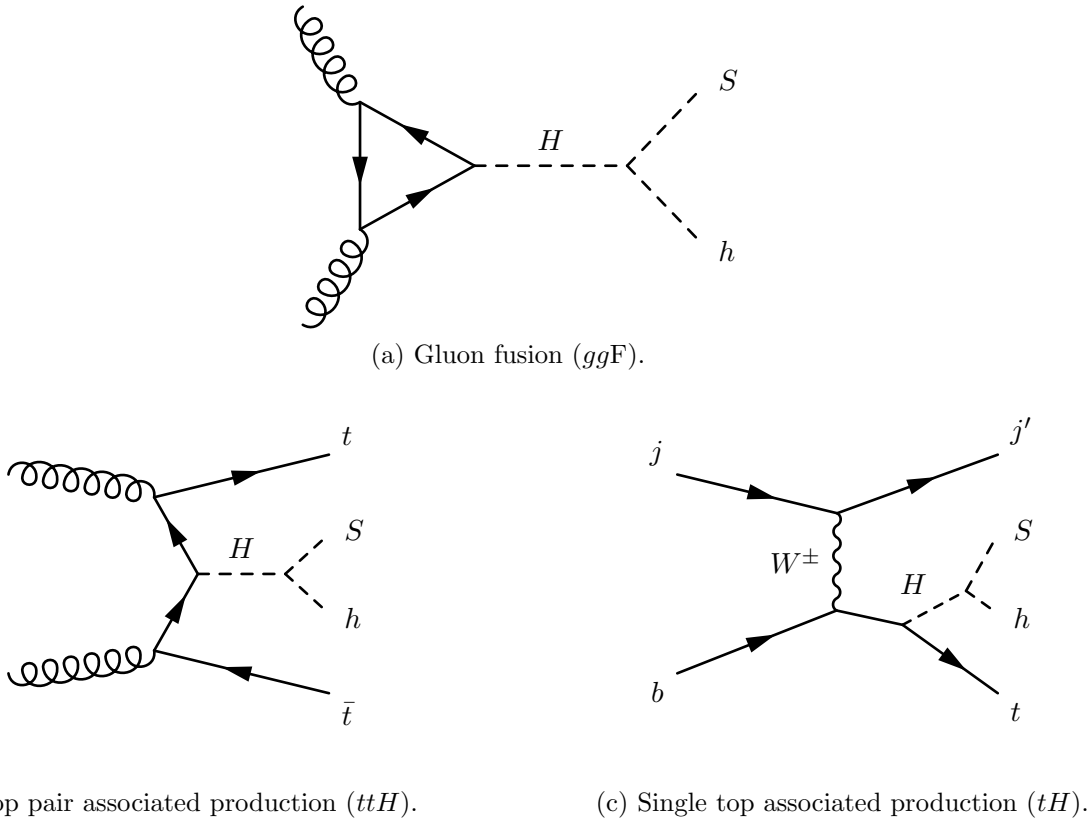

The first term in \Cref{eqn:H_production} allows for the gluon fusion (\ggf) production mode of $H$.
Due to the squaring of the matrix element in width calculations, production cross sections involving this Yukawa coupling are scaled by $\bgs$.
The value of \bgs is used as a free parameter in fits to the data.
We have set $\beta_{_V}=0$, such that the coupling of $H$ to pairs of the weak vector bosons is significantly small; the associated production of $H$ with the weak vector bosons and vector boson fusion (VBF) are negligible production modes.
The dominant production mode of $H$ is therefore \ggf, while both single ($tH$) and double ($ttH$) top associated production of $H$ are also non-negligible.
While single top associated production of a Higgs-like boson is usually suppressed due to interference, the implicit assumption of a significantly small $H$-$V$-$V$ coupling allows for a sizeable $tH$ production cross section~\cite{Farina:2012xp}.
It has been shown in previous studies~\cite{vonBuddenbrock:2017gvy,vonBuddenbrock:2015ema} that the $tH$ cross section is enhanced to being approximately that of the $ttH$ cross section.
The representative Feynman diagrams for the production modes of $H$ are shown in \Cref{fig:feynman_diagrams}.

The $S$ boson, on the other hand, is assumed not to be produced directly but rather through the decay of $H$.
In principle, it is possible to include $S$ as a singlet scalar that has interactions with $H$ and the SM Higgs boson $h$.
Doing this would allow the $H$ to produce $S$ bosons through the $H\to SS$ and $Sh$ decay modes.
Here we assume the $H\to Sh$ decay mode to have a 100\% BR (also shown in \Cref{fig:feynman_diagrams}).
These assumptions are all achieved by introducing the following effective interaction Lagrangians.
Firstly, $S$ is given a vacuum expectation value and couples to the scalar sector:
\begin{equation}
{\cal L}_{HhS} = -\frac{1}{2}~v\Big[\lambda_{_{hhS}} hhS + \lambda_{_{hSS}} hSS +
\lambda_{_{HHS}} HHS   + \lambda_{_{HSS}} HSS + \lambda_{_{HhS}} HhS\Big], \label{eqn:HS_coupling}
\end{equation}
where the couplings are fixed to ensure that the $H\to Sh$ BR is 100\%. In order to reduce the parameter space $S$ is assumed to be a SM Higgs-like scalar. 

\section{Anatomy of the anomalies}
\label{sec:anatomy}

All fits to the LHC data here were performed using a template-based method of performing fits based on maximizing a profile likelihood ratio.
The SM components of the fits are always taken directly from the published experimental distributions, along with their associated systematic uncertainties, including those that affect normalization and shape. The BSM component is always constructed using a single mass point ($m_H=270$~GeV and $m_S=150$~GeV) and therefore has only one degree of freedom under the assumptions stated in \Cref{sec:model}.
The single degree of freedom is $\bgs$, which maps directly to the normalisation of the BSM signal with respect to the SM Higgs-like production cross section of $H$. It is very important to note that the boson masses were not tuned. They were fixed to values obtained with an analysis obtained from different data sets corresponding to Run 1 data, as reported in Ref.~\cite{vonBuddenbrock:2017gvy}. Further, the choice of final states under study here was made based on the predictions of Ref.~\cite{vonBuddenbrock:2016rmr}. In contrast to  searching for excesses in a wide span of the phase-space, the excesses identified here are related to a prediction.

The statistical likelihood function $L\left(\bgs~|~\theta\right)$ is constructed as the product of Poisson probabilities for each bin and in each considered measurement. Systematic uncertainties are incorporated as additional constraint factors in the likelihood, which vary according to their associated nuisance parameters $\theta$. The best-fit value of the parameter of interest $\bgs$ is identified as the minimum of $-2\log\lambda\left(\bgs\right)$, where a deviation of one unit in this quantity is equivalent to a $1\sigma$ deviation from the best-fit point of the parameter of interest.
Since the value $\bgs=0$ corresponds to the SM-only hypothesis (the \textit{null} hypothesis), the significance of each fit is calculated as:
\begin{equation}
    Z=\sqrt{-2\log\lambda\left(0\right)}.
    \label{eqn:significance}
\end{equation}

The fits include searches for the SM production of top quarks decaying to opposite-sign (OS) lepton pairs, searches for Higgs boson production in leptonic final states and BSM searches for the production of same-sign (SS) lepton pairs, to name a few.
Many of these searches involve either a signal or dominant background component that contains top quarks in the final state.
Therefore, the results are often always dependent on the number of $b$-jets produced with the leptons ($e,\mu$).

\begin{table}
    \centering
       \caption{A list of the ATLAS and CMS experimental results pertaining to final states with multiple leptons that are considered here.
    For each result, a simple baseline selection is shown.}
    \begin{tabular}{lll}
    \br
        \multicolumn{1}{c}{\textbf{Data set}} & \multicolumn{1}{c}{\textbf{Reference}} & \multicolumn{1}{c}{\textbf{Selection}}  \\
        \mr
        ATLAS Run 1 & ATLAS-EXOT-2013-16\hfill\cite{Aad:2015gdg} & SS $\ell\ell$ and $\ell\ell\ell$ + $b$-jets \\
        ATLAS Run 1 & ATLAS-TOPQ-2015-02\hfill\cite{Aaboud:2017ujq} & OS $e\mu$ + $b$-jets \\
        CMS Run 2 & CMS-PAS-HIG-17-005\hfill\cite{CMS-PAS-HIG-17-005} & SS $e\mu$, $\mu\mu$ and $\ell\ell\ell$ + $b$-jets \\
        CMS Run 2 & CMS-TOP-17-018\hfill\cite{Sirunyan:2018lcp} & OS $e\mu$ \\
        CMS Run 2 & CMS-PAS-SMP-18-002\hfill\cite{CMS-PAS-SMP-18-002} & $\ell\ell\ell+\MET$ ($WZ$) \\
        ATLAS Run 2 & ATLAS-EXOT-2016-16\hfill\cite{Aaboud:2018xpj} & SS $\ell\ell$ and $\ell\ell\ell$ + $b$-jets \\
        ATLAS Run 2 & ATLAS-CONF-2018-027\hfill\cite{ATLAS-CONF-2018-027} & OS $e\mu$ + $b$-jets \\
        ATLAS Run 2 & ATLAS-CONF-2018-034~\hfill\cite{ATLAS-CONF-2018-034} & $\ell\ell\ell+\MET$ ($WZ$) \\
        \br
    \end{tabular}
 
    \label{tab:results_list}
\end{table}

\begin{table}
    \centering
       \caption{A summary of the SM+BSM fit results for each measurement considered here, along with the result of their combination. DFOS stands for different flavor and opposite sign, in relation to di-leptons.}
    \begin{tabular}{lcc} \br
    
        \multicolumn{1}{c}{\textbf{Selection}} & \multicolumn{1}{c}{\textbf{Best-fit} \bgs} & \multicolumn{1}{c}{\textbf{Significance}}  \\
        \mr
        ATLAS Run 1 SS leptons + $b$-jets & $6.51\pm2.99$ & 2.37$\sigma$  \\
        ATLAS Run 1 DFOS di-lepton + $b$-jets & $4.09\pm1.37$ & 2.99$\sigma$ \\
        ATLAS Run 2 SS leptons + $b$-jets & $2.22\pm1.19$ & 2.01$\sigma$ \\
        CMS Run 2 SS leptons + $b$-jets & $1.41\pm0.80$ & 1.75$\sigma$ \\
        CMS Run 2 DFOS di-lepton & $2.79\pm0.52$ & 5.45$\sigma$ \\
        ATLAS Run 2 DFOS di-lepton + $b$-jets & $5.42\pm1.28$ & 4.06$\sigma$ \\
        CMS Run 2 tri-lepton + $\MET$ & $9.70\pm3.88$ & 2.36$\sigma$ \\
        ATLAS Run 2 tri-lepton + $\MET$ & $9.05\pm3.35$ & 2.52$\sigma$ \\
        \mr
        Combination & $2.92\pm0.35$ & 8.04$\sigma$ \\
        \br
    \end{tabular}
    \label{tab:combination_list}
\end{table}

\begin{figure}[t]
\begin{center}
\begin{minipage}{14pc}
\includegraphics[angle=90,width=16pc]{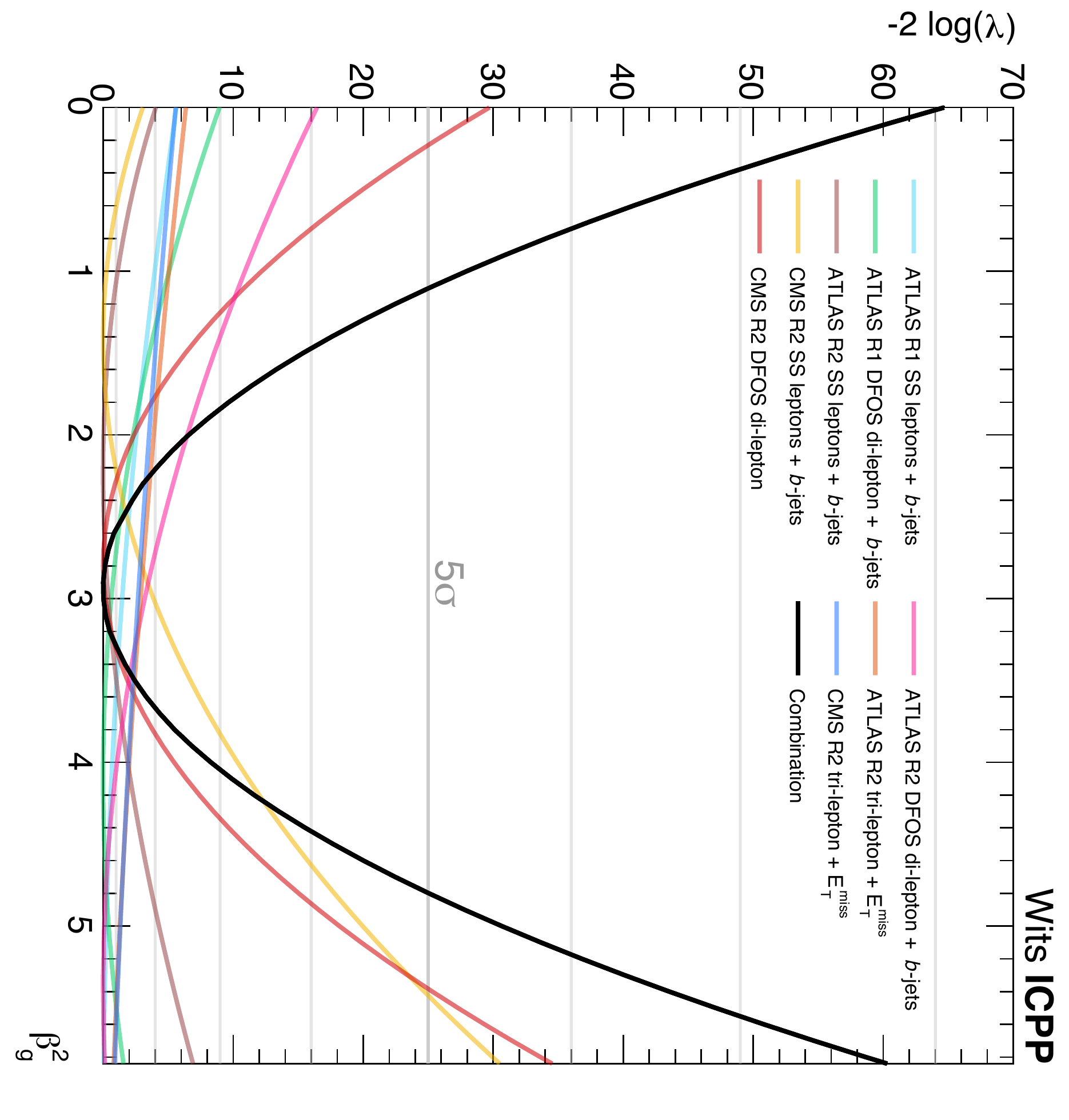}
\caption{\label{fig:plr} The  profile likelihood ratios for each of the individual fit results and their combination~\cite{vonBuddenbrock:2019ajh}.
    The significance of a result is calculated as the square root of the point which intersects the $y$-axis.}
\end{minipage}\hspace{2cm}%
\begin{minipage}{14pc} \vspace{-0.5cm}\hspace{-1cm}
\includegraphics[width=17pc]{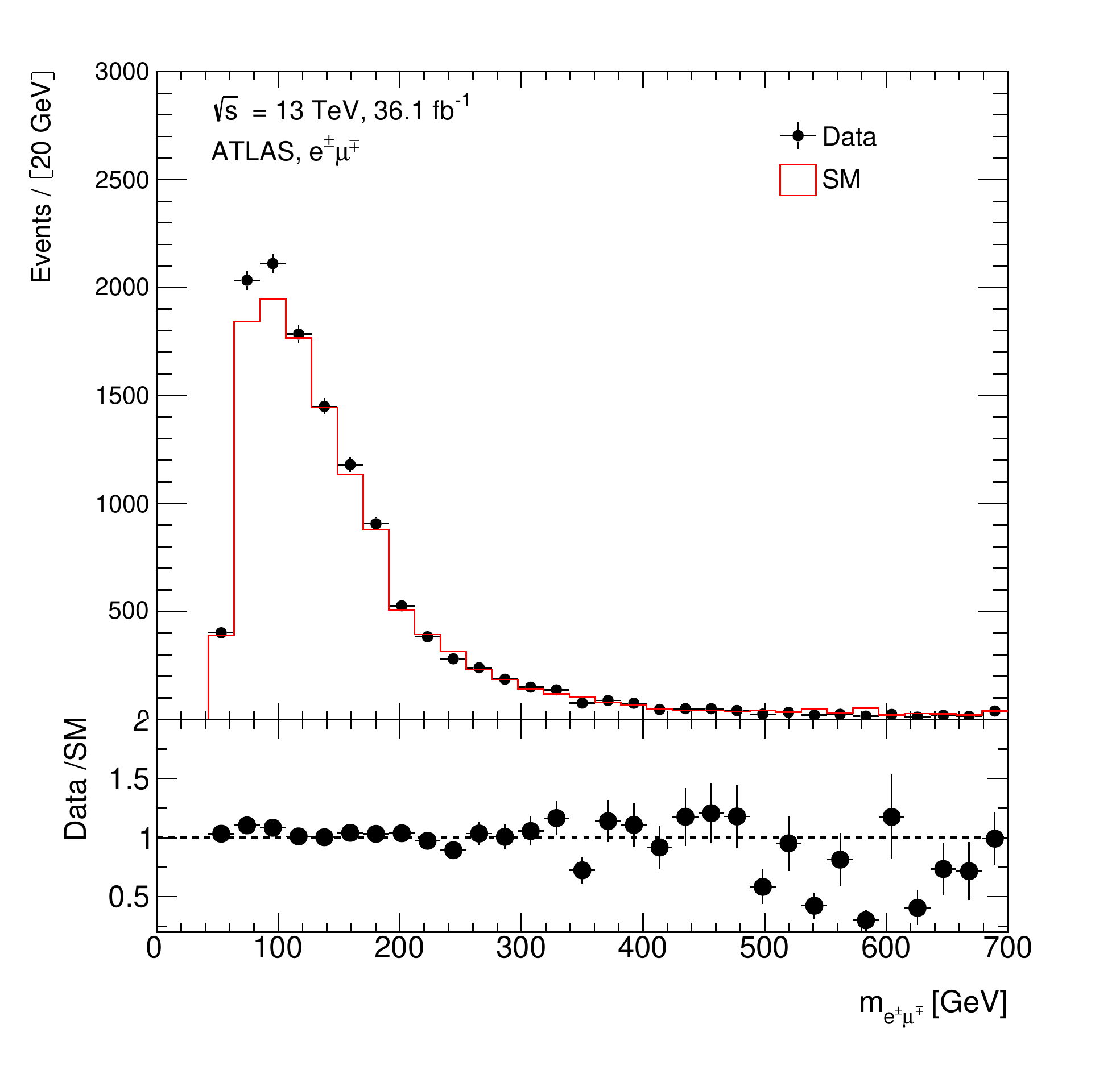}
\caption{\label{fig:mll} Dilepton invariant spectrum in events with a full hadronic jet veto after the application of NNLO QCD and NLO EW corrections  (see text). 
}
\end{minipage} 
\end{center}
\end{figure}

The ensemble of results considered in this article is shown in \Cref{tab:results_list}.
The majority of results come from the Run 2 data sets. Each of the results studied in this article make use of a profile likelihood ratio to constrain the single fit parameter \bgs under an SM+BSM hypothesis.
With these profile likelihood ratios constructed as a function of \bgs, it is relatively straightforward to perform a simultaneous fit on all of the results considered and therefore make a combination of the independent data sets under the SM+BSM hypothesis.
The combined profile likelihood is constructed by multiplying the profile likelihood ratios for each individual measurement.
Then, the best-fit value of \bgs and significance can be calculated similarly to the individual results.
Doing so constrains the parameter \bgs to the value $2.92\pm0.35$, which corresponds to a significance of $Z=8.04\sigma$ in favour of the SM+BSM hypothesis over the SM-only hypothesis.
A summary of all the individual fit results, as well as the combination, can be seen in \Cref{tab:combination_list}.
In addition to this, each of the individual profile likelihood ratios are shown in \Cref{fig:plr}, with the combined case shown in black. The statistical significance reported here is obtained with a simplified model, which imperfectly describes the deviations of the data with respect to the SM. As such, the significance reported here is a conservative estimate. It is remarkable that a simplified model, such as the one used here, is able to accommodate such a wide range of discrepancies. These  include the production of two opposite sign leptons, same sign leptons and three leptons for which the production cross-sections are vastly different. 

Recent results reported by the ATLAS collaboration confirm with more statistics the anomalies described above in di-lepton final states with a full hadronic jet veto~\cite{Aaboud:2019nkz}, where the dominant SM process is the non-resonant production of $W$ pairs. \Cref{fig:mll} displays the di-lepton invariant mass in $e\mu$ events with a full hadronic jet veto after the application of the aforementioned corrections. Here QCD NNLO corrections to $q\overline{q}\rightarrow W^+W^-$ production~\cite{Gehrmann:2014fva,Grazzini:2016ctr,Hamilton:2016bfu,Re:2018vac}, QCD NLO corrections to non-resonant $gg\rightarrow W^+W^-$~\cite{Caola:2015rqy} and EW NLO corrections~\cite{Biedermann:2016guo} have been applied. The SM MC has been normalized to the data with $m_{\ell\ell}>110$\,GeV. The discrepancy re-emerges here with $m_{\ell\ell}<100$\,GeV, as predicted in Refs.~\cite{vonBuddenbrock:2016rmr,vonBuddenbrock:2017gvy}, showing similar features compared to the discrepancies in di-leptom final states with $b$-jets documented in Ref.~\cite{vonBuddenbrock:2019ajh}. The deviation seen in \Cref{fig:mll} is not included in the excess reported in \Cref{fig:plr}, and neither were those already identified with Run 1 data in Ref.~\cite{vonBuddenbrock:2017gvy}. 

\Cref{tab:anatomy} summarizes the final states studied here, including basic characteristics and the corresponding dominant SM process. The anomalies described here appear in final states and corners of the phase-space where different SM processes dominate. This important feature renders the possibility of explaining the anomalies by MC mismodelling rather improbable. 

\begin{table}
    \centering
       \caption{A succinct summary of the characteristics of the multi-lepton anomalies studied here. }
    \begin{tabular}{lcc} 
    \br
        \multicolumn{1}{c}{\textbf{Final State}} & \multicolumn{1}{c}{\textbf{Characteristic}} & \multicolumn{1}{c}{\textbf{Dominant SM process}}  \\
        \mr
        $\ell^+\ell^-$ + jets, $b$-jets & $m_{\ell\ell}<100$\,GeV & $t\overline{t}, Wt$ \\
        $\ell^+\ell^-$ + jet veto & $m_{\ell\ell}<100$\,GeV & $W^+W^-$ \\
        $\ell^\pm\ell^\pm$ + $b$-jets & Moderate $H_T$& $t\overline{t}V, (V=Z,W^{\pm})$ \\
        $\ell^\pm\ell^\mp\ell^\pm$ + $b$-jets & Moderate $H_T$ & $t\overline{t}V, (V=Z,W^{\pm})$ \\
        $Z(\rightarrow \ell^+\ell^-)+\ell^\pm$ & $p_{TZ}<100$\,GeV & $W^{\pm}Z$ \\
        \br
    \end{tabular}
    \label{tab:anatomy}
\end{table}

\section{Conclusions}
\label{sec:conclusions}

A number of predictions were made in Refs.~\cite{vonBuddenbrock:2015ema,vonBuddenbrock:2016rmr} pertaining to the anomalous production of multiple leptons at high energy proton-proton collisions. These would be connected with a heavy boson with a mass around the EW scale decaying predominantly into a SM Higgs boson and a singlet scalar. Discrepancies in multi-lepton final states were reported with Run 1 data in Refs~\cite{vonBuddenbrock:2015ema,vonBuddenbrock:2017gvy} have now become statistically compelling with the available Run 2 data~\cite{vonBuddenbrock:2019ajh}. These include the production of opposite-sign, same-sign and three leptons with and without $b$-jets. Discrepancies emerge in final states and corners of the phase-space where different SM processes dominate, indicating that the potential mismodeling of a particular SM process is unlikely to explain them. The yields of the anomalies and their kinematic characteristics are remarkably well described by a simple ansatz, where $H\rightarrow Sh$ is produced via gluon-gluon fusion and in association with top quarks. 

\section*{References}
\bibliography{references}

\end{document}